\documentclass[article]{aa} 

\usepackage{graphicx}
\usepackage{amssymb}
\usepackage{natbib}
\begin{document}
   \title{The paucity of  globular clusters around  the field elliptical NGC\,7507
   }

   \titlerunning{The paucity of  globular clusters around  the field elliptical NGC\,7507}

   \author{J. P. Caso
          \inst{1}
          \and
          T. Richtler
          \inst{2}
          \and 
          L.P. Bassino
          \inst{1}
          \and
          R. Salinas
          \inst{3}
          \and 
          R. R. Lane
          \inst{2}
       \and 
         A. Romanowsky
        \inst{4}
          }

\authorrunning{ J. P. Caso et. al.}

   \institute{Facultad de Ciencias
Astron\'omicas y Geof\'isicas de la Universidad Nacional de La Plata;
Consejo Nacional de Investigaciones Cient\'ificas y T\'ecnicas; and
Instituto de Astrof\'isica de La Plata (CCT La Plata-CONICET-UNLP),
Argentina 
         \and
Departamento de Astronom\'{\i}a,
Universidad de Concepci\'on,
Concepci\'on, Chile
         \and
        Finnish Centre for  Astronomy with ESO,
University of Turku,
V\"ais\"al\"antie 20,
FI-21500 Piikki\"o,
Finland     
          \and 
             Department of Physics \& Astronomy,
             San Jos\'e State University, San Jos\'e, California, USA}

   \date{Received ...; accepted ...}

 
  \abstract
  {There is strong evidence that
  globular cluster systems (GCSs) of massive galaxies are largely
  assembled by infall/accretion processes. Therefore, we expect the
  GCSs of isolated elliptical galaxies to be poor. Although not
  completely isolated, NGC 7507 is a massive field elliptical galaxy
  with an apparently very low dark matter content.}
   {We determine the richness, the colour distribution, and the
     structural properties of the GCS of NGC 7507.}
   {We perform wide-field Washington photometry with data obtained
     with the MOSAIC II camera at the 4m-Blanco telescope, CTIO.}
   {The GCS is very poor with {\bf $S_N \approx 0.6$}. We identify three
     subpopulations with peaks at $(C-T1)$ colours of 1.21, 1.42, and
     1.72. The bluest population may represent the old, metal-poor
     component. This interpretation is supported by its shallow
     density profile. The red population is more concentrated,
     resembling the galaxy light. The intermediate-colour population is
     strongly peaked in colour and we interpret this population as the signature
     of a starburst, whose age depends on the metallicity, but should
     be quite old, since no signatures of a merger are
     identifiable. In addition, we find a main sequence in the stellar
     foreground population, which we attribute to the Sagittarius
     dwarf tidal stream. 
     }
  {The extrordinarily  poor GCS of NGC 7507, a  massive elliptical galaxy, is an
    illustration of how important the environmental conditions are for
    producing rich GCSs.}


   \maketitle
%

\section{Introduction}
\label{sec:intro}
Elliptical galaxies are predominantly found in groups and clusters of
galaxies. Isolated elliptical galaxies, on the other hand, are rare
and, therefore, mostly distant. 
The term
``isolated ellipticals'' goes back to photographic times
\citep{kar73}, when elliptical galaxies were understood as being
featureless, with faint features not easily visible. Recently,
\citet{tal09} showed that the majority of isolated ellipticals in
their sample showed tidal features, which could indicate previous
merger activity.

Ellipticals in clusters have accreted a significant part of their
material since $z$=2 \citep{vdo10}, see also \citet{jim11}. The possibility of accreting
material is plausibly weaker for isolated galaxies, so this could
constitute a major difference between isolated elliptical galaxies and
ellipticals in clusters, and could be manifest in their globular
cluster systems (see \citealt{ric12a} for a recent review).
Not much is known about globular cluster systems (GCSs) of isolated
elliptical galaxies.  That the  richness of a GCS is related to its local galaxy density,
is, however, not a new insight \citep{wes93}.
 A compilation of GCS properties, including the
environmental density of host galaxies, is given by \cite{spi08}. The
few isolated ellipticals appearing in that study have relatively low
$S_N$-values \footnote{The specific frequency $S_N$ is defined as
$S_N = N_{GC} \times 10^{0.4 (M_V +15)}$, $N_{GC}$ beeing the total
number of globular clusters of a GC system and $M_V$ the absolute
magnitude of the host galaxy.}.  
Also recent work on the isolated ellipticals NGC 5812 and NGC 3585
reveal very low $S_N$-values \citep{lan12}.
However, this is not always the case as the sample of 
galaxies in low-density environments
 by \citep{cho12} shows. In this context,
NGC 7507 is a particularly interesting object. Its morphology is that
of a ``real'' elliptical galaxy (E0) without clear signs of tidal
distortions. Dark matter is not necessarily required to explain its
kinematics \citep{sal12}, however, some dark matter may be present, in
conjunction with a radial anisotropy of stellar orbits.  Even in the
case where some dark matter exists, the total dark matter content is
much lower than expected for an elliptical galaxy of this mass
\citep{nie10}. NGC 7507 has a spiral galaxy companion (NGC 7513) at a
projected distance of 18\arcmin, and is, therefore, not {\it truly}
isolated. Here we investigate its globular cluster system using
wide-field Washington photometry. This photometric system has proven
particularly useful for investigations of GCS in numerous studies
\citep{dir03a,dir03b,dir05,bas06a,bas06b}.
We adopt a distance of 23.2\,Mpc (m-M=31.83). At this distance,
1\arcsec corresponds to 112.5\,pc, and the absolute magnitude within
projected 50 kpc is $M_R = -22.64$ according to the photometric model of
\citet{sal12}, corresponding to a stellar mass of $2\times10^{11} M_\odot$.

\section{Observations and Reductions}

\begin{figure}[htbp]
\begin{center}
 \includegraphics[width=90mm]{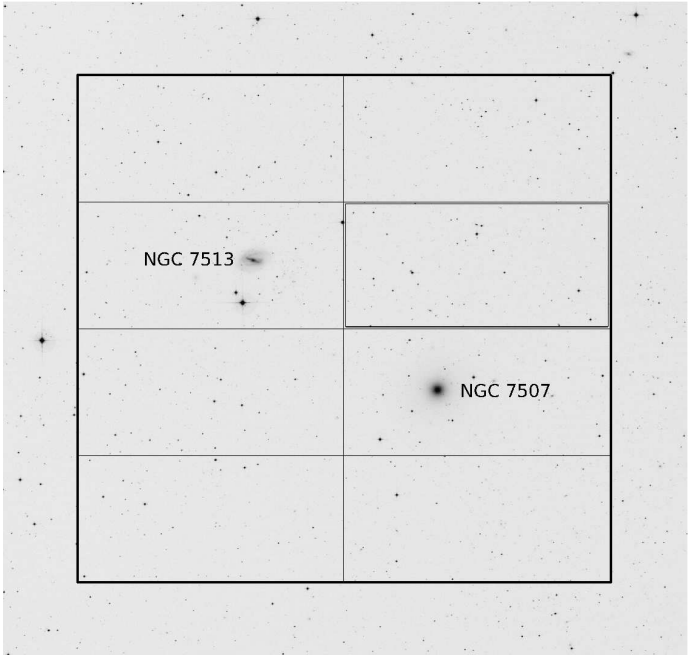}
 \caption{Our MOSAIC II field around NGC 7507, overlaid onto a DSS image. The field is $36\arcmin
   \times 36\arcmin$. North is up, east to the left. The companion galaxy NGC 7513 is seen toward the north-east.  The eight individual CCDs of the MOSAIC II camera are indicated. The chip\,\#3 is identified by a double line.}
\label{fig:mosaic}
\end{center}
\end{figure}

The observations were performed with the MOSAIC\,II camera (8 CCDs mosaic imager, 16 amplifiers) 
mounted at the prime focus of the 4-m Blanco telescope at the Cerro Tololo Inter-American 
Observatory (CTIO, Chile), during 2005 August 5 -- 6. One pixel of the MOSAIC wide-field 
camera subtends 0.27\,arcsec on the sky (http://www.ctio.noao.edu/mosaic), which corresponds 
to a field of $36 \times 36~{\rm arcmin}^2$ or approximately $230\,{\rm kpc} \times 230\,{\rm kpc}$ at the 
adopted distance of NGC\,7507. The details of the observations have been given in \citet{sal12}, 
where some of these images were used. 

The field is shown in Figure \ref{fig:mosaic} and was observed with Kron-Cousins $R$ and Washington 
$C$ filters. We selected the $R$ 
filter instead of the original Washington $T_1$ as \citet{gei96a} has shown that the Kron-Cousins 
$R$ filter is more efficient than $T_1$ (see \citealt{sal12} for a more thorough explanation). 
After the calibration, $C$ and $R$ instrumental magnitudes were transformed directly  into  $C$ and $T_1$ standard magnitudes.

The data were dithered to fill in the gaps between the eight individual MOSAIC chips 
(see Figure \ref{fig:mosaic} to identify the location of the different chips on the field). 
Originally, four images in $R$ with exposure times of 720\,s each, and seven images in $C$ 
with exposures of 1800\,s each were obtained (shorter exposures of 60\,s in $R$ and of 300\,s 
in $C$ were also obtained to avoid saturation at the galaxy core). However, according to CTIO 
reports, the first night was ``useful'' and only the second one was ``photometric''. In order 
to improve the photometry, the calibration has been revised for the present paper. Also, to 
obtain the final combined images, we will use all of the long exposure $R$ images, but just five 
out of the seven $C$ ones, i.e. those observed in the second night with better seeing.

The tasks in the MSCRED package within IRAF were applied to reduce the MOSAIC data, 
following some of the guides given at http://iraf.noao.edu/projects/ccdmosaic/. 
Due to the large field of the MOSAIC images, the pixel scale across the CCDs is variable.  
As a consequence, the brightness of point sources may differ up to 4\% between image centre and
corners.
The MSCRED package contains tasks to correct for this effect. 
For instance, the 
MSCCMATCH task uses a reference celestial catalogue (an astrometric catalog) to correct 
for any difference between the reference coordinates and the observed ones, i.e. updates 
the world coordinate system (WCS) of the image to the WCS of the reference system. 
In a later stage, the MSCIMATCH task takes into account any possible differences in the 
intensity scale between the chips, homogenizing them according to the intensities of a 
reference image.   
The final combined $R$ image resulted with remaining sensitivity variations up to 1.4\%, 
and the final $C$ image up to 3.0\%, calculated from peak-to-peak over the whole images. 
In these final images, the seeing is $1.1\arcsec$ and $1.3\arcsec$ in the $R$ and $C$ frames, 
respectively.

The chip\,\#3 (Fig.\,\ref{fig:mosaic}) 
caused problems during the final stacking of the images, 
performed with the MSCSTACK task. Those regions where chip\,\#3 
overlapped with other CCDs, by the dithering procedure, caused 
multiple images to be created. Masking out the respective part 
of chip\,\#3 solved the problem, but introduced some inhomogeneity.

\section{Photometry}

In order to facilitate point source detection, the extended galaxy
light was subtracted using a ring median filter \citep{sec95}, with 
an inner radius of $8.0\arcsec$ and an outer radius of $9.3\arcsec$. 
These values were selected according to our previous experience with 
MOSAIC images \citep[e.g.][]{dir03a,dir03b}, and it was checked that 
the median filter had not affected the point source photometry.
An initial selection of point sources was obtained
with the software SExtractor \citep{ber96}, using a gaussian filter on
the $C$ image. We considered every group of at least five connected
pixels above a threshold of $1.5\sigma$ as a positive detection
(DETECT\_MINAREA and DETECT\_TRESH parameters,
respectively). Effective radii of classical GCs do not usually exceed
a few parsecs \citep[e.g.][]{har09}, which implies that, at the
distance to NGC\,7507, GCs
are seen as point-sources on MOSAIC\,II images. To select point sources
we used the star/galaxy classifier from SExtractor, through the
CLASS$\_$STAR parameter, that takes values close to one for point-sources 
and close to zero for extended sources. All objects with CLASS$\_$STAR$<0.4$ were
rejected and about 14000 point-sources were detected on the whole MOSAIC field.

The photometry was performed using Daophot/IRAF in the usual manner. A
second-order variable PSF was built for each final $R$ and $C$ image,
employing about one hundred bright stars well distributed over the
field. The final point source selection was based on the $\chi$ and
sharpness parameters from the ALLSTAR task.

\begin{figure}
 \includegraphics[width=90mm]{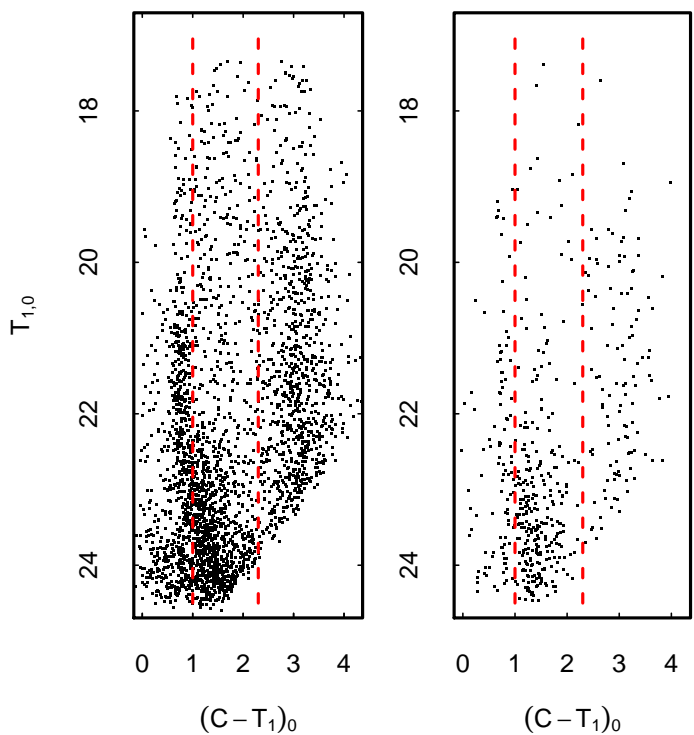}
 \caption{Colour-magnitude diagram for the point sources in the MOSAIC
   II field. The left panel shows the entire field, while the right
   panel shows only those objects closer than 7\arcmin\ to the center 
   of NGC 7507. The
   red dashed lines indicate the colour range for old GCs. It is clear
   that the GCS is quite poor. Striking is the main sequence feature
   which probably belongs to the Sagittarius stream. See
   the Appendix for further discussion.}
 \label{fig:cmdmos}
\end{figure}

As stated above, the calibration performed by \citet{sal12} has been 
improved for the present paper. Three fields of standard stars
selected from the list of \citet{gei96a} were used to derive the
photometric calibration. The revised set of calibration equations
(only $C$ differs from the those used by \citealt{sal12}) is as
follows:
\begin{eqnarray*}
{T_1} & = & {R}_\mathrm{inst}+(0.89\pm0.01)-(0.23\pm0.01)X_\mathrm{R}\\
        & & {} +(0.016\pm0.003)({C}-{T_1}),\\
{C} & = & {C}_\mathrm{inst}+(1.54\pm0.01)-(0.66\pm0.01)X_\mathrm{C}\\
        & & {} +(0.103\pm0.003)({C}-{T_1})
\end{eqnarray*}
where ${R}_\mathrm{inst}$ and ${C}_\mathrm{inst}$ are the instrumental
magnitudes and $X_\mathrm{R}$ and $X_\mathrm{C}$ are the airmasses,
respectively. 

The aperture corrections and their corresponding errors, 
computed from the stars used to build each PSF, are $-0.33\pm0.02$
and  $-0.32\pm0.02$ for the $C$ and $R$ photometry, respectively.
The $(B-V)$ reddening was calculated from the \citet{sch98} maps. 
The colour excess in Washington photometric system was
 adopted  as
$E_{(C-T_1)} = 1.97 \times E_{(B-V)}$ \citep{har77} and for the $T_1$ 
absorption we used the relation $A_R/A_V = 0.75$ \citep{rie85}. Thus, 
the foreground reddening and absorption corrections applied were 
$E_{(C-T_1)} = 0.09$ and $A_{T1} =0.12$, respectively.

\subsection{Completeness estimation}

In order to estimate the completeness of our final sample, we added to each
final $R$ and $C$ image 1000 artificial stars, equally distributed over the field, covering the magnitude and colour intervals expected
for GCs. This process was 
repeated 10 times, reaching a total sample of 10000 artificial stars. The
photometry of these new images was done in the same way as the original one.

\begin{table}[!hts]
\begin{minipage}{90mm}
\begin{center}
\caption{Completeness up to $7'$ from the NGC 7507 centre} 
\label{tab_compl}
\begin{tabular}{@{}cc@{}}
\hline
\\
\multicolumn{1}{c}{$T_1$}&\multicolumn{1}{c}{Completeness}\\
\multicolumn{1}{c}{$[mag]$}&\multicolumn{1}{c}{$\%$}\\

\hline
\\
$19.5-20.0$&$0.95$\\
$20.0-20.5$&$0.92$\\
$20.5-21.0$&$0.92$\\
$21.0-21.5$&$0.85$\\
$21.5-22.0$&$0.81$\\
$22.0-22.5$&$0.76$\\
$22.5-23.0$&$0.64$\\
$23.0-23.5$&$0.62$\\
$23.5-24.0$&$0.58$\\
$24.0-24.5$&$0.30$\\
\hline
\end{tabular} 
\end{center} 
\end{minipage}
\end{table}

The completeness function around the galaxy indicates that the $75\%$ 
completeness is achieved at $T_1\approx 22.3$, and the $40\%$ at 
$T_1\approx 24$. We adopt $T_1=24$ as our magnitude limit. In 
Table\,\ref{tab_compl} we indicate the completeness for bins of
$0.5\,mag$.

\section{The globular cluster system}

\subsection{Selection of globular cluster candidates}

We selected as GC candidates those point-sources with $1<(C-T_1)_0<2.3$ and
$T_{1,0}>20$. 
This colour range corresponds to the entire metallicity range
for old clusters and is similar to that used in the literature 
for previous Washington photometric studies of GCs 
\citep[e.g.][]{gei96b,dir03a,bas06a,ric12b}. Figure \ref{fig:cmdmos}
shows the colour-magnitude diagram (CMD) for all point sources in the
entire field (left panel) and for point sources closer than
7\arcmin\ to the galaxy center (right panel). We expect the universal 
turn-over-magnitude of old cluster systems
at about $m_R \approx 24.4$ (see Sect.\ref{sec:numbers}). It is clear that this is
not a rich GCS. 
 Red dashed lines indicate
the $(C-T_1)_0$ range for the GC candidates. The left panel does not show anything 
of a GCS and even for the radius selected sample, the existence of a GCS is not obvious.

 A striking feature of the
CMD is the vertical sequence of objects bluer than $(C-T_1)_0 = 1$.
The distribution of these point sources turns out to be uniform over
the field.

These sources are accounted for in the general background subtraction. 
The adopted background region is $\approx 12'$ away from NGC\,7507, and its area 
is almost $700\,{\rm arcmin}^2$. That is, the background is far from the projected GCS. 
In the Appendix, we show that this 
``strip'' is probably a main sequence of a population belonging to the tidal 
stream of the Sagittarius dwarf galaxy \citep{iba94}.

\subsection{Colour distribution}
\label{sec:colour}
\begin{figure}
 \includegraphics[width=90mm]{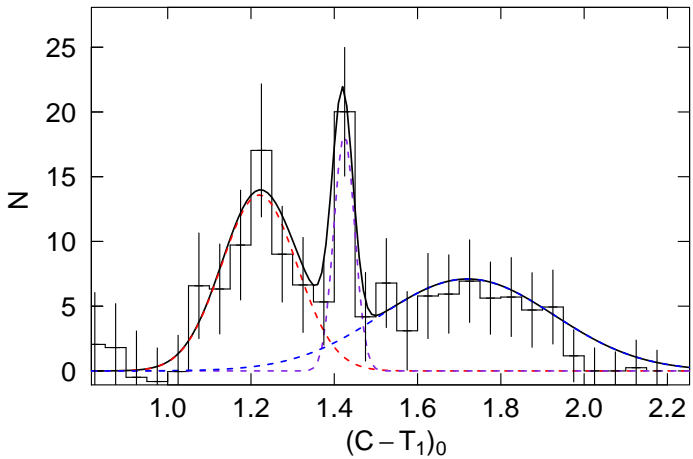}
 \caption{Background corrected colour distribution for GC candidates in
   the radial regime $0.5' < R < 7'$. A sum of three gaussians has
   been fitted and is plotted as solid black lines, while each
   individual gaussian is plotted as dashed lines. We interpret the
   sharp peak at $(C-T_1)=1.4$ as the signature of a starburst.}
 \label{fig:dcol}
\end{figure}

Figure \ref{fig:dcol} shows the background-corrected colour
distribution for GC candidates brighter than $T_1 = 24$ in the radial
regime $0.5'<R<7'$ (raw and background counts are listed in 
Table\,\ref{tab_dcol}). The outer radial limit is argued for in Section
\ref{sec:radial}, the inner limit avoids strong incompleteness due to
the galaxy brightness. The bin width is $0.05$.  

The main observation from Figure \ref{fig:dcol} is that the GC colour
distribution differs significantly from the ``normal'' distributions
of GCs in giant elliptical galaxies. Most striking is the sharp peak
in the colour range $(C-T_1)_0 \sim1.40-1.45$.

\begin{table}[!hts]
\begin{minipage}{90mm}
\begin{center}
\caption{ The first column indicates the bins in which the colour range was 
divided. The second column shows the raw counts for the
colour distribution of GC candidates in the radial regime $0.5' < R < 7'$.
The third one shows the background counts, normalized with the ratio between
the areas of the science and background regions.} 
\label{tab_dcol}
\begin{tabular}{@{}ccc@{}}
\hline
\\
\multicolumn{1}{c}{$(C-T_1)_0$}&\multicolumn{1}{c}{Raw}&\multicolumn{1}{c}{Background}\\
\multicolumn{1}{c}{$[mag]$}&\multicolumn{1}{c}{}&\multicolumn{1}{c}{}\\

\hline
\\
$0.90-0.95$&$8$&$8$\\
$0.95-1.00$&$4$&$5$\\
$1.00-1.05$&$5$&$5$\\
$1.05-1.10$&$12$&$6$\\
$1.10-1.15$&$10$&$4$\\
$1.15-1.20$&$15$&$5$\\
$1.20-1.25$&$23$&$6$\\
$1.25-1.30$&$12$&$3$\\
$1.30-1.35$&$11$&$4$\\
$1.35-1.40$&$9$&$4$\\
$1.40-1.45$&$23$&$3$\\
$1.45-1.50$&$9$&$5$\\
$1.50-1.55$&$9$&$3$\\
$1.55-1.60$&$7$&$4$\\
$1.60-1.65$&$9$&$3$\\
$1.65-1.70$&$8$&$2$\\
$1.70-1.75$&$9$&$2$\\
$1.75-1.80$&$7$&$1$\\
$1.80-1.85$&$7$&$2$\\
$1.85-1.90$&$7$&$2$\\
$1.90-1.95$&$7$&$2$\\
$1.95-2.00$&$3$&$2$\\
$2.00-2.05$&$1$&$2$\\
$2.05-2.10$&$1$&$2$\\
\hline
\end{tabular} 
\end{center} 
\end{minipage}
\end{table}

We fit the colour distribution by a sum of three gaussians and obtain
as mean peak colours $(C-T_1)_0=1.21\pm0.02$, $(C-T_1)_0=1.42\pm0.02$
and $(C-T_1)_0=1.72\pm0.04$, respectively. Comparing these values with
those available in the literature
(e.g. \citealt{dir03b,bas06a,bas06b}), the colour of the
redder gaussian is typical of metal-rich GC populations. On the other
hand, metal-poor GC populations usually present peak colours in the
range $1.25 < (C-T_1)_0 < 1.36$, slightly redder than that of the blue
population in NGC 7507. However, in Figure \ref{fig:dcol} the blue GC
candidate distribution seems to be confined to $(C-T_1)_0 > 1$, which
is expected for old GCs. The dispersion for the bluest and reddest
gaussians are, $\sigma_B = 0.08\pm 0.01$ and $\sigma_R = 0.20\pm
0.06$, respectively, which are comparable to the values found in other
galaxies. Remarkable is the dispersion of the intermediate-colour
peak. Its dispersion has an upper limit of 0.025 mag. This is of the
order of the photometric uncertainty and the intrinsic dispersion is,
therefore, likely even smaller.

The authors are aware of only one other example of an intermediate-age
starburst being visible in the colour distribution of a GCS, that
associated with NGC 1316 \citep{ric12b}.
Figure 2 by \citet{ric12a} shows that spectroscopically confirmed 
intermediate-age GCs from NGC 1316 are found in a very tight $(C-T_1)_0$ 
range in the CMD.

Thus, we have the three GC groups $1 < (C-T_1)_0 < 1.35$, $1.37 <
(C-T_1)_0 < 1.47$ and $1.47 < (C-T_1)_0 < 2.2$ which we label as
``blue'', ``intermediate'' and ``red'' samples, respectively. Once the
background has been subtracted, each group represents 41, 20 and
$39\,\%$, of the total number of GC candidates, respectively. The 
appearance of the colour distribution does not depend strongly on the 
width of the colour bins. It does depend on the brightness in the sense 
that the red clusters are mainly faint clusters which supports the 
interpretation that 
 the younger and 
brighter clusters are found among the blue objects.

\begin{figure}
 \includegraphics[angle=270,width=90mm]{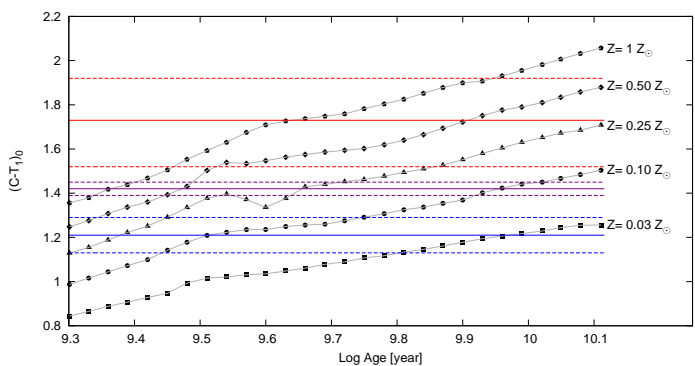}
 \caption{Theoretical models of single stellar populations by
   \citet{bre12} for five different metallicities (filled
   symbols). Blue, violet and red solid lines indicate the mean colours
   of the three GC groups, while dashed lines indicate their
   dispersions.}
 \label{fig:mod}
\end{figure}

Figure \ref{fig:mod} compares the mean colours of the three groups with
theoretical models of single stellar populations by \citet{bre12},
using their web-based tool (http://stev.oapd.inaf.it/cgi-bin/cmd). The
models are plotted for five different metallicities (filled points),
spanning a wide range. Blue, violet and red solid lines indicate the
mean colours of the ``blue'', ``intermediate'' and ``red'' populations,
respectively, while dashed lines indicate their dispersions.  If the
age estimation for NGC 7507 itself by \citet{sal12} is adopted (i.e.
8-10\,Gyr), the metallicities of the ``blue'' and ``red'' samples are
consistent with those of metal-poor and metal-rich old GCs
\citep[e.g.][]{bro06}.

If the GCs from the intermediate sample have solar metallicity, Fig.\ref{fig:mod} 
indicates 
an age of about 2-3 Gyr.
 However, a significant field population of this age
does not seem to be present (see the Discussion for further remarks).

\subsection{Projected spatial and radial distributions}
\label{sec:radial}
\begin{figure*}
 \includegraphics[width=180mm]{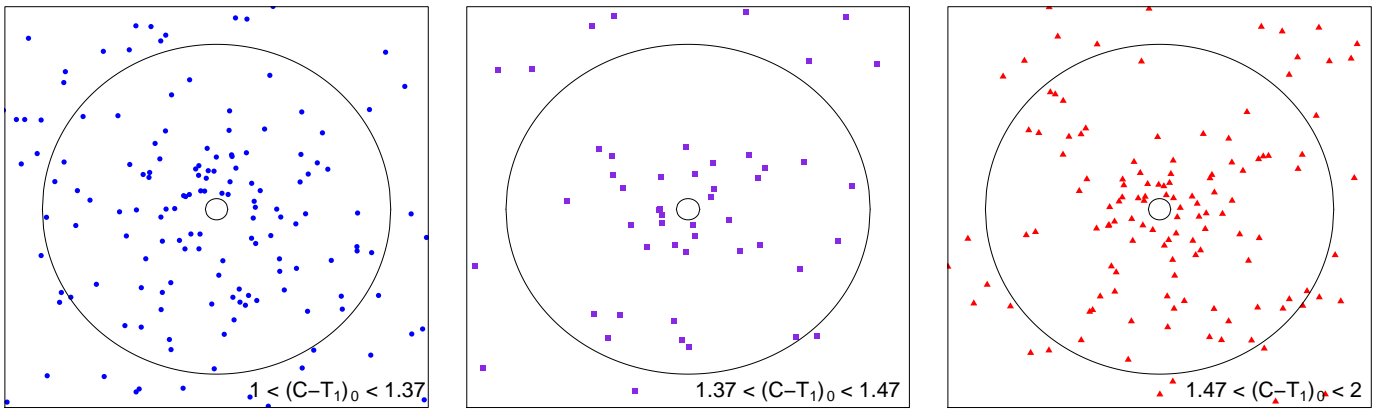}
 \caption{Projected spatial distribution of GC candidates, separated
   in three different colour bins. The circles plotted in each panel
   refer to the selected radial regime for the analysis
   (i.e. $0.5'<R<7'$). Each figure spans $15.75' \times 15.75'$. 
North is up, east to the left.}
 \label{fig:esp}
\end{figure*}

Figure \ref{fig:esp} shows the projected spatial distribution of GC
candidates in the radial regime $0.5'<R<7'$, separated by the three
different colour intervals defined previously. The circles in each
panel represent the limits of the chosen radial regime. Comparing the
distribution of the blue (left panel) and red sample (right panel),
the latter seems to be slightly more concentrated towards the galaxy
center. This is consistent with the usual radial behaviour of
metal-poor and metal-rich GCs in elliptical galaxies. The spatial
distribution of the intermediate sample (middle panel) seems to prefer
smaller radii than the blue sample and one might also see a slight
preference for the east-west direction, which, however, is weakly defined.

\begin{figure}
 \includegraphics[width=90mm]{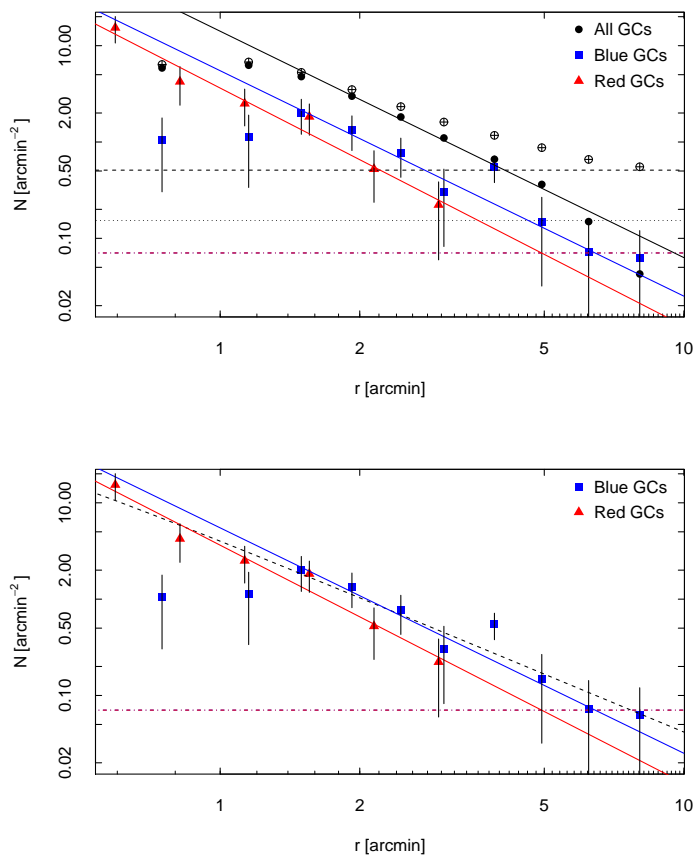}
 \caption{{\bf Upper panel :} Raw radial distribution for all the GC
   candidates (open circles). Background corrected distributions for
   GC candidates with $1 < (C-T_1)_0 < 2.3$ (filled black circles),
   and for the blue (blue squares) and red (red triangles)
   samples. The solid lines represent power-laws (see text) for the respective samples corresponding
   by colour. The dashed  line indicates the background level, while
   the dotted one refers to the $30\,\%$ of the background
   level. The dash-dotted line is the $30\,\%$ of the background level
   for the ``blue'' and ``red'' samples, which are indistinguishable. 
{\bf Lower panel:}  Dashed
   line indicates the Hubble-Reynolds profile of NGC\,7507 by
   \citet{sal12}. }
 \label{fig:rad}
\end{figure}

The projected radial distribution of the GC candidates are shown in
Figure \ref{fig:rad}. Panel (a) shows the raw distribution for all GC
candidates (open circles), and the background corrected distributions
for the complete colour range (filled black circles) as well as for the
``blue'' (blue squares) and ``red'' (red triangles) samples. The
horizontal dashed line indicates the background level, while the
dotted line refers to $30\%$ of the background level. We propose this
limit to define the GCS extension. The dash-dotted line
corresponds to $30\%$ of the background level in the colour ranges of
the ``blue'' and ``red'' samples.

We estimate the extension of the NGC 7507 GCS to approximately $7'$
(i.e. 47.25\,kpc).  It is not possible to fit a uniform power-law to
the distribution of the blue sample because of the two deviating
innermost points. These counts cannot be low due to completeness,
because the red clusters should have been at least equally
affected. In the radial range 1.5\arcmin-5\arcmin\
the logarithmic density slopes of the blue and the red sample are
indistinguishable. Their values are $-2.4\pm0.15$ and $-2.5\pm0.2$,
respectively.

In the lower panel we compare the radial distributions of the blue and
red samples with the light profile of NGC 7507,
which is the Hubble-Reynolds profile given by \citet{sal12}.  An
arbitrary scaling has been applied for comparison purposes.
All slopes are similar, but the deficiency of blue clusters at small
radii is striking.

 The small size of the ``intermediate'' sample prohibits obtaining a
 radial density profile.  Instead, we statistically subtract the
 background contribution in the three samples. For this purpose, we
 introduce 7 annuli with a width of 1\arcmin\ each to cover the full
 radial range. From each we erase random background objects according
 to the background density.
 The mean distance of objects to the galaxy center was then obtained
 for each sample.  This procedure was repeated one hundred
 times. Finally, the average of the mean distances and their errors
 were $3.3\pm0.2\arcmin$, $2.8\pm0.3\arcmin$ and $2.6\pm0.2\arcmin$
 for the blue, intermediate and red sample, respectively.
A Kolmogorov-Smirnov test \citep{kol33,smi48} was used to compare the
 radial distribution of the  clean ``intermediate'' sample with the others for each iteration. In
all cases, the probability that blue and intermediate samples could be
described by the same distribution is almost null. A similar result
was obtained when blue and red samples were compared. For the
comparison between the intermediate and red samples, a 50\% of
probability was obtained.

\subsection{Dust in NGC 7507}
Morphologically, NGC 7507 has, except for a central dust lane \citep{spa85}, a
smooth appearance without obvious shells or tidal structures (the
feature reported by \citealt{tal09} had not been confirmed by
\citealt{sal12}).  \citet{sal12} showed the  central dust structures in
the colour map in short exposure images. The colours are too red to
resemble a stellar population and are plausibly caused by reddening.
 Here we describe our long exposure colour map,
which shows that there is also dust distributed on a larger scale.

\begin{figure}
 \includegraphics[angle=90,width=90mm]{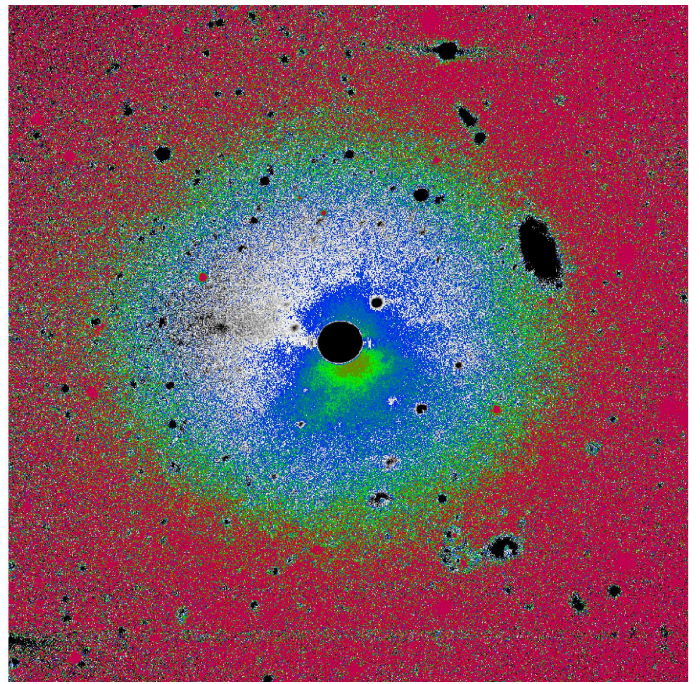}
 \caption{ $4' \times 4'$ NGC 7507 colour map. North is up, East to the
   left. The colour palette represents the range
   $1.85<(C-T_1)<2.2$ from blue to red (beyond a radius of about 1\arcmin\,  the colours are not longer
   trustworthy). We interpret the redder parts (indicated with blue and green colours) as being due to dust, which has been detected before \citep{tem04}.}
 \label{fig:map}
\end{figure}

Figure \ref{fig:map} displays a $4'\times4'$ NGC 7507 colour map.
The colour palette spans the range $1.85 < (C-T_1)_0 < 2.2$. It is
clear that the colour distribution is not radially symmetric. 
The redder parts, which plausibly are caused by  reddening due to dust, extend to
about 1\arcmin\ (6.7 kpc) to the south-west. 
The existence of dust in an otherwise  morphologically undisturbed galaxy could  be evidence of a
past merger and is perhaps related to the existence of the intermediate population of GCs. 
See the discussion for more remarks.

\subsection{Total numbers and specific frequency}
\label{sec:numbers}

In this Section
we  estimate the number of members 
for the NGC\,7507 GCS  and its specific frequency $S_N$. 
 For the radial interval 
$0.5\arcmin<R<7\arcmin$, we count the background corrected objects in 
bins of $0.5$ mag, down to $T_1=24$ mag, which is our $40\%$ completeness 
limit. To correct this for the missing central region we adopt
for the innermost 0.5\arcmin\ the same surface density as for the
first bin, resulting in a correction of 5 objects (i.e., $\approx 4\%$
of the $0.5\arcmin<R<7\arcmin$ background corrected sample). This 
criterion was applied by other authors in GCs studies (e.g.,
\citealt{bas06b}), assuming that the radial distribution of GCs might be 
flatter in the very inner region of a galaxy (e.g., \citealt{els98}).
Finally, we correct the number of objects in each bin by its corresponding
incompleteness, arriving at approximately 210 GC candidates. 
We still need to calculate the number of GCs fainter than $T_1=24$ mag. 
It is expected that an old GC population presents a 
gaussian luminosity function with a turn-over magnitude (TOM) in the 
$V$-band of $M_{\rm TOM}\approx-7.4$ (e.g. \citealt{ric03,jor07})
 at our adopted distance at an apparent magnitude
of $T_1\approx 24.4$. The mean $(V-R)$ colour of the GCS is not drastically
different from 0.4, so that our limiting magnitude marks roughly the
TOM. This allows us to double the calculated number of GCs brighter than 
$T_1=24$ mag to obtain a total population of approximately 420 GC candidates.
 
Our photometric model of NGC 7507 \citep{sal12} gives an absolute
brightness of $M_{\rm R}=-22.64$ (extinction of 0.12 mag included)
within 7\arcmin\ (47.25 kpc) or $M_{\rm V}=-22.16$ (adopting
$(V-R)=0.6$ from \citealt{pru98}). With this brightness, $S_N \approx
0.6$. This value changes with adopted distance, however, it will stay
far below what is known for elliptical galaxies in cluster
environments, even adopting its closest estimate.

\section{Discussion}

\subsection{The total number of GCs and the environment}
The number of GCs around NGC 7507 is only about one tenth of the
number of GCs one normally finds around cluster ellipticals of
comparable magnitude (e.g. \citealt{bro06,ric12a}).

Illustrative
is the comparison with NGC 4636, an elliptical giant galaxy in the
outskirts of the Virgo cluster: while the stellar masses of the
galaxies are comparable, NGC 4636 has about 6000 GCs \citep{dir05}.
Since the galaxy density around NGC 4636 is not extremely high, this shows that
the present-day environmental density is not a very strict parameter for determining
the richness of a GCS. 

Still the  literature is not entirely clear in this respect. In
the compilation by \citet{spi08}, the few isolated elliptical galaxies
indeed have low $S_N$-values, but in the recent work of \citet{cho12}
on the GCSs of  early-type galaxies in low-density environments, some appear with higher
$S_N$-values, NGC 1172 even with $S_N$ = 9.2$\pm$4.4. 
(however, we cannot reproduce this value with the data given in their paper). The conclusion of \citet{cho12} that the host galaxy mass is the
primary parameter in determining the properties of a GCS, cannot be backed-up with
NGC 7507. 
We moreover remark that our $S_N$-value cannot be directly compared
with those of \citet{cho12}, which are systematically too high with respect
to our value.
The use of the RC3-magnitudes for NGC 7507 for brightness and colour would have
meant an absolute brightness of $M_V = -21.5$, about 0.7 magnitudes
fainter than the brightness of our photometric model. 

In addition,  \citet{cho12}   only use the density
parameter of  \citet{tul88} to define a low-density environment. None of the Cho et al. galaxies appear in 
catalogues of {\it isolated} ellipticals 
(e.g. \citealt{red04,smi04,sto04,col06}).
 In particular, NGC 1172 appears as a member of the "NGC
1209 group" according to \citet{san78}.

\subsection{Subpopulations and a possible starburst}
Bimodality is the normal concept for the substructure of the optical
colour distribution of giant ellipticals.  However, it might disappear
when using infrared colours \cite[][]{chi12}. Its significance is under
discussion (e.g. \citealt{ric06,yoo06,ric12a}) due to the relation 
between colours and metallicities being unclear. Although we identify three
subpopulations, the intermediate population has a special character in
that it may have its origin in a starburst younger than the main
population. Our blue population does have a bluer peak colour than
known from giant ellipticals \citep{bas06a}.

If we consider the accretion of dwarf galaxies as the dominant process
in the assembly of the metal-poor component of GCSs
\citep{cot98,hil99,pen08,sch10} there might be a straight-forward
explanation. In a poor environment with less GC donators to accrete,
we expect, according to the shape of the galaxy luminosity function, a
bias toward faint and, therefore, metal-poor galaxies. As a
consequence, the accreted GCs are also more metal-poor and the
resulting colour peak may be bluer and the width of the Gaussian
smaller.

For radii larger than $1.5'$, no difference of the slopes between red 
and blue clusters is discernable. However, blue clusters can be followed to 
larger radii than red clusters. For radii smaller than $1.5'$ there is a clear
deficiency of blue GCs. This fits 
to a scenario in which the red ({\it bona fide} metal rich) clusters trace the 
formation of the bulge population, while the blue ({\it bona fide} metal-poor) 
clusters have mostly been accreted through the accretion of their former host 
galaxies (see \citealt{ric12a} for a recent account on the formation of globular cluster systems). This supports the general picture and adds confidence that 
the blue objects really are the old, metal-poor GC population.

Even if spectroscopic confirmation must have the last word, the peak
at $(C-T_1)_0\sim 1.4$ seems to indicate a population with a
homogeneous age and/or metallicity. The natural origin would be a star
burst, triggered by a close encounter, perhaps with NGC 7513, or a
merger/infall of a gas-rich galaxy. Another example of a GCS showing
peaks in the colour distributions related to starbursts is NGC 1316,
where the merger properties are obvious \citep{ric12b}. This is not
the case in NGC 7507, where shells or tidal tails cannot be
identified.  Assigning an age to the peak colour, one must know the
metallicity. At solar metallicity, the age would be about 2-3 Gyr.
The existence of a significant field population of this age is not
plausible, given the relatively red colour of the galaxy. More
plausible, therefore, is the hypothesis of a lower metallicity and
higher age (see Fig.\ref{fig:mod}). The infall of a gas- rich dwarf galaxy would
be a valid scenario, which could also be responsible for the amount of
dust found in NGC 7507.
\citet{tem04} quotes
a mass of $130\times10^{4} M_\odot$ (their two-component model), which
is low compared with star-burst galaxies, but definitely shows the
existence of dust and the possibility of internal reddening.

\subsection{Dark matter - GC connection}
An interesting suggestion has been made by \citet{spi08}.  From their
compilation of dark halo masses and properties of GCSs, they found
that ``GCs formed in direct proportion to the halo mass''. If we
replace ``formed'' by ``appeared'' then we avoid the underlying
suggestion that all GCs form in their respective host galaxies. An
important perception of elliptical galaxy evolution is that since
$z$=2, the stellar masses of elliptical galaxies in clusters have
grown by a factor of about 4 through accretion processes
\citep{vdo10}. Accretion in the $\Lambda$CDM context means accretion
of galaxies, probably with a large mass range, including their GCSs
and dark matter haloes. If the rich GCSs of central galaxies grow by
accretion, it is plausible that in poor environments with less
material to accrete, the GCSs are poor and the dark matter content is
low. \citet{sal12} has shown that the kinematics of NGC 7507 can be
explained without any requirement for dark matter.
However, a small dark matter halo may be permitted under some radial
anisotropy increasing outwards. The stellar mass of NGC 7507 within 50
kpc is about $2\times10^{11} M_\odot$, adopting the model of
\citet{sal12}. Adopting \citet{nie10} as reference for the assignment
of dark matter to stellar matter in $\Lambda$CDM simulations, one
would expect a dark halo mass of the order $10^{13} M_\odot$. This is
clearly not supported by the data. A halo mass of about $6\times
10^{11} M_\odot$ could be marginally accomodated.

 We can evaluate how NGC 7507 fits in the relation between
total mass $M_{180}$ (which is the sum of baryonic and dark mass within a volume
with an average density of 180 times the background density) and the mass in GCs, given by \citet{spi09}:
$$M_{180} = 14125 \times M_{GCS}$$
Adopting 260 as the number of clusters  and, following \citet{spi09},
 $4\times 10^5 M_\odot$ as the average GC mass, we have $1.04 \times 10^8 
M_\odot$ as the total mass in GCs. 
Therefore, one expects $1.4 \times 10^{12} M_\odot$ as $M_{180}$. Using 
$2 \times 10^{11} M_\odot$ (\ref{sec:intro}) for the stellar mass, we see that
the expected mass is higher by a factor of about 2 than the mass suggested by the
observations.

\section{Summary and conclusions}
We investigated the globular cluster system of NGC 7507, a massive
elliptical field galaxy.

The globular cluster system of NGC 7507 is very poor, fitting more to
a spiral galaxy than to a giant elliptical galaxy. For the specific
frequency, we derive the extraordinarily low value of $S_N = 0.3$,
which is in line with other galaxies in low-density environments.
The colour distribution shows the blue and red peak, commonly described
by the term ``bimodality'' in the GCSs of elliptical galaxies. The red
peak has a similar colour to that found for the GCSs of other host
galaxies. However, the blue peak has a colour somewhat bluer than
normally found in richer cluster systems, agreeing with other studies
of isolated ellipticals \citep{cho12}. A further well-known feature is
the shallower number density profile of the blue clusters. A
peculiarity is a sharp peak at $(C-T_1)=1.4$, making up 20\% of the
cluster population.

We suggest the following interpretation:\\Due to its relatively
isolated location, only a few accreted dwarf galaxies donated their
GCSs to NGC 7507, limiting the metal-poor component. The peak at
$(C-T1)=1.4$ is the signature of a starburst with an $apriori$ unknown
age. The most plausible hypothesis is the infall/merger of a gas-rich
dwarf galaxy. This could have happened 7-8 Gyr ago and would be
consistent with a younger age, independently indicated by the low
$M/L_R$-ratio of 3 \citep{sal12}.
 
Furthermore, as the dark matter content of
NGC 7507 is low, one may see it as an example supporting the relation
of \citet{spi08}, where GCs appear roughly in proportion to the dark
halo mass.

\begin{acknowledgements}
JPC and LPB acknowledge financial support
by grants from Universidad Nacional de La Plata and Consejo Nacional
de Investigaciones Cient\'ificas y T\'ecnicas CONICET (Argentina).
TR acknowledges financial support from the Chilean Center for Astrophysics,
FONDAP Nr. 15010003,  from FONDECYT project Nr. 1100620, and
from the BASAL Centro de Astrofisica y Tecnologias
Afines (CATA) PFB-06/2007.

R.R.L. gratefully acknowledges financial support from FONDECYT, project
No. 3130403. 
\end{acknowledgements}

\appendix
\section[]{Sagittarius dwarf galaxy tidal tail in the foreground of NGC 7507}

It is apparent that there is an unusual sequence in the CMD of point
sources from the NGC 7507 field (Figure \ref{fig:cmdmos}), with
$T_{1,0}\gtrsim20.0$ and $0.7\lesssim (C-T_1)_0\lesssim1.8$. This feature is
reminiscent of CMD sequences ascribed to both the Monoceros
Ring, an apparently extra-Galactic ring of tidal debris encircling the
Disc \cite[e.g.][]{con07,con08,con12}, and to detections of the
Sagittarius (Sgr) dwarf \cite[][]{iba94} tidal stream
\cite[e.g.][]{mat96,maj99}. Furthermore, the latitude of this field is
too high $(b\sim-68)$, and has a magnitude range that is much too
small, to be attributed to the Galactic disc, or any other known
Galactic substructure.

\begin{figure}
 \includegraphics[width=90mm]{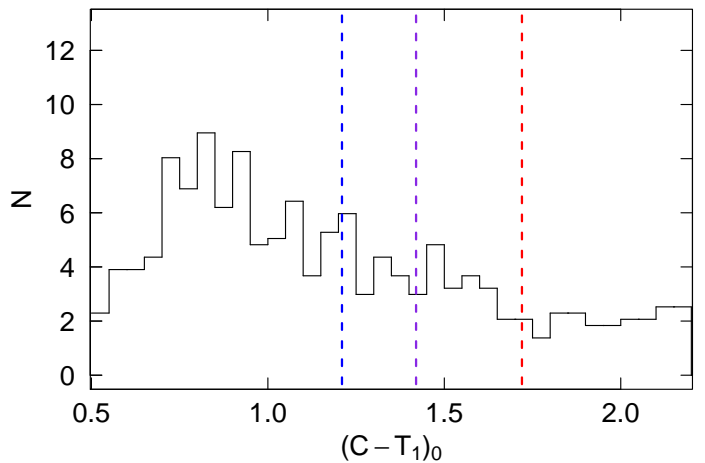}
 \caption{Colour distribution of the point sources in the background 
region, scaled down to the size of the galaxy field. 
The dashed lines indicate the mean colours of the three gaussians 
fitted to the GC colour distribution in Section \ref{sec:colour}, 
$(C-T_1)_0=1.21$, $(C-T_1)_0=1.42$ and $(C-T_1)_0=1.72$, respectively.}
 \label{fig:backg}
\end{figure}

Interestingly, this field is located at
(RA,dec)~$\sim(23^h12^m,-28^\circ32'')$, a location well-placed to
include stellar members of the tidal stream from the Sgr dwarf. The
colour of the vertical component of the CMD sequence ($[C-T_1]\sim0.7$,
Figure \ref{fig:cmdmos}) correlates well with the $(V-I)$ colour of
the same part of the sequence \cite[${[}V-I{]}\sim0.8$;][]{mat96,maj99},
assuming a metalliticy of $Z=0.08$ (colour transformations derived
following \citealt{bre12} with corrections by
\citealt{gir10}). Because of the obvious similarities between the CMD
sequence shown in Figure \ref{fig:cmdmos}, the location of the field
and the color of the sequence, we attribute this feature to a
detection of the Sgr tidal stream in the foreground of NGC 7507.
Figure \ref{fig:backg} shows the colour distribution of the 
point sources from the background region, scaled down to
the size of the galaxy field. According to the CMD
shown in Figure \ref{fig:cmdmos}, the bins with the highest counts
are located in the colour range $0.7\lesssim (C-T_1)_0\lesssim1$. For
larger colour values, redder bins present lower count values. The dashed 
lines indicate the mean colours of the three gaussians fitted to the GC 
colour distribution in Section \ref{sec:colour}. The counts in the
range $1\lesssim (C-T_1)_0\lesssim2.2$ imply that the three peaks in the
GC colour distribution would persist even if the contamination was 
increased by a $50\,\%$.

 A comparison with recent results from Pan-STARRS1 \citep{sla13} shows
 that  the line-of-sight toward NGC 7507 crosses the bright arm in the southern
 part of the Sgr stream. 

Note, the CMD sequence broadens toward dimmer magnitudes. Because
photometry is intrinsically more uncertain at fainter magnitudes, this
broadening is most likely due to uncertainties in the photometry, and,
therefore, does not reflect a stream with a large systemic distance
spread between stellar stream members.

\label{lastpage}

\bibliographystyle{aa}
\bibliography{biblio}

\begin{thebibliography}{57}
\expandafter\ifx\csname natexlab\endcsname\relax\def\natexlab#1{#1}\fi

\bibitem[{{Bassino} {et~al.}(2006{\natexlab{a}}){Bassino}, {Faifer}, {Forte},
  {Dirsch}, {Richtler}, {Geisler}, \& {Schuberth}}]{bas06a}
{Bassino}, L.~P., {Faifer}, F.~R., {Forte}, J.~C., {et~al.} 2006{\natexlab{a}},
  \aap, 451, 789

\bibitem[{{Bassino} {et~al.}(2006{\natexlab{b}}){Bassino}, {Richtler}, \&
  {Dirsch}}]{bas06b}
{Bassino}, L.~P., {Richtler}, T., \& {Dirsch}, B. 2006{\natexlab{b}}, \mnras,
  367, 156

\bibitem[{{Bertin} \& {Arnouts}(1996)}]{ber96}
{Bertin}, E. \& {Arnouts}, S. 1996, \aaps, 117, 393

\bibitem[{{Bressan} {et~al.}(2012){Bressan}, {Marigo}, {Girardi}, {Salasnich},
  {Dal Cero}, {Rubele}, \& {Nanni}}]{bre12}
{Bressan}, A., {Marigo}, P., {Girardi}, L., {et~al.} 2012, \mnras, 427, 127

\bibitem[{{Brodie} \& {Strader}(2006)}]{bro06}
{Brodie}, J.~P. \& {Strader}, J. 2006, \araa, 44, 193

\bibitem[{{Chies-Santos} {et~al.}(2012){Chies-Santos}, {Larsen}, {Cantiello},
  {Strader}, {Kuntschner}, {Wehner}, \& {Brodie}}]{chi12}
{Chies-Santos}, A.~L., {Larsen}, S.~S., {Cantiello}, M., {et~al.} 2012, \aap,
  539, A54

\bibitem[{{Cho} {et~al.}(2012){Cho}, {Sharples}, {Blakeslee}, {Zepf}, {Kundu},
  {Kim}, \& {Yoon}}]{cho12}
{Cho}, J., {Sharples}, R.~M., {Blakeslee}, J.~P., {et~al.} 2012, \mnras, 422,
  3591

\bibitem[{{Collobert} {et~al.}(2006){Collobert}, {Sarzi}, {Davies},
  {Kuntschner}, \& {Colless}}]{col06}
{Collobert}, M., {Sarzi}, M., {Davies}, R.~L., {Kuntschner}, H., \& {Colless},
  M. 2006, \mnras, 370, 1213

\bibitem[{{Conn} {et~al.}(2007){Conn}, {Lane}, {Lewis}, {Gil-Merino}, {Irwin},
  {Ibata}, {Martin}, {Bellazzini}, {Sharp}, {Tuntsov}, \& {Ferguson}}]{con07}
{Conn}, B.~C., {Lane}, R.~R., {Lewis}, G.~F., {et~al.} 2007, \mnras, 376, 939

\bibitem[{{Conn} {et~al.}(2008){Conn}, {Lane}, {Lewis}, {Irwin}, {Ibata},
  {Martin}, {Bellazzini}, \& {Tuntsov}}]{con08}
{Conn}, B.~C., {Lane}, R.~R., {Lewis}, G.~F., {et~al.} 2008, \mnras, 390, 1388

\bibitem[{{Conn} {et~al.}(2012){Conn}, {No{\"e}l}, {Rix}, {Lane}, {Lewis},
  {Irwin}, {Martin}, {Ibata}, {Dolphin}, \& {Chapman}}]{con12}
{Conn}, B.~C., {No{\"e}l}, N.~E.~D., {Rix}, H.-W., {et~al.} 2012, \apj, 754,
  101

\bibitem[{{Cote} {et~al.}(1998){Cote}, {Marzke}, \& {West}}]{cot98}
{Cote}, P., {Marzke}, R.~O., \& {West}, M.~J. 1998, \apj, 501, 554

\bibitem[{{Dirsch} {et~al.}(2003{\natexlab{a}}){Dirsch}, {Richtler}, \&
  {Bassino}}]{dir03b}
{Dirsch}, B., {Richtler}, T., \& {Bassino}, L.~P. 2003{\natexlab{a}}, \aap,
  408, 929

\bibitem[{{Dirsch} {et~al.}(2003{\natexlab{b}}){Dirsch}, {Richtler}, {Geisler},
  {Forte}, {Bassino}, \& {Gieren}}]{dir03a}
{Dirsch}, B., {Richtler}, T., {Geisler}, D., {et~al.} 2003{\natexlab{b}}, \aj,
  125, 1908

\bibitem[{{Dirsch} {et~al.}(2005){Dirsch}, {Schuberth}, \& {Richtler}}]{dir05}
{Dirsch}, B., {Schuberth}, Y., \& {Richtler}, T. 2005, \aap, 433, 43

\bibitem[{{Elson} {et~al.}(1998){Elson}, {Grillmair}, {Forbes}, {Rabban},
  {Williger}, \& {Brodie}}]{els98}
{Elson}, R.~A.~W., {Grillmair}, C.~J., {Forbes}, D.~A., {et~al.} 1998, \mnras,
  295, 240

\bibitem[{{Geisler}(1996)}]{gei96a}
{Geisler}, D. 1996, \aj, 111, 480

\bibitem[{{Geisler} {et~al.}(1996){Geisler}, {Lee}, \& {Kim}}]{gei96b}
{Geisler}, D., {Lee}, M.~G., \& {Kim}, E. 1996, \aj, 111, 1529

\bibitem[{{Girardi} {et~al.}(2010){Girardi}, {Williams}, {Gilbert},
  {Rosenfield}, {Dalcanton}, {Marigo}, {Boyer}, {Dolphin}, {Weisz},
  {Melbourne}, {Olsen}, {Seth}, \& {Skillman}}]{gir10}
{Girardi}, L., {Williams}, B.~F., {Gilbert}, K.~M., {et~al.} 2010, \apj, 724,
  1030

\bibitem[{{Harris} \& {Canterna}(1977)}]{har77}
{Harris}, H.~C. \& {Canterna}, R. 1977, \aj, 82, 798

\bibitem[{{Harris}(2009)}]{har09}
{Harris}, W.~E. 2009, \apj, 699, 254

\bibitem[{{Hilker} {et~al.}(1999){Hilker}, {Infante}, \& {Richtler}}]{hil99}
{Hilker}, M., {Infante}, L., \& {Richtler}, T. 1999, \aaps, 138, 55

\bibitem[{{Ibata} {et~al.}(1994){Ibata}, {Gilmore}, \& {Irwin}}]{iba94}
{Ibata}, R.~A., {Gilmore}, G., \& {Irwin}, M.~J. 1994, \nat, 370, 194

\bibitem[{{Jim{\'e}nez} {et~al.}(2011){Jim{\'e}nez}, {Cora}, {Bassino},
  {Tecce}, \& {Smith Castelli}}]{jim11}
{Jim{\'e}nez}, N., {Cora}, S.~A., {Bassino}, L.~P., {Tecce}, T.~E., \& {Smith
  Castelli}, A.~V. 2011, \mnras, 417, 785

\bibitem[{{Jord{\'a}n} {et~al.}(2007){Jord{\'a}n}, {McLaughlin},
  {C{\^o}t{\'e}}, {Ferrarese}, {Peng}, {Mei}, {Villegas}, {Merritt}, {Tonry},
  \& {West}}]{jor07}
{Jord{\'a}n}, A., {McLaughlin}, D.~E., {C{\^o}t{\'e}}, P., {et~al.} 2007,
  \apjs, 171, 101

\bibitem[{{Karachentseva}(1973)}]{kar73}
{Karachentseva}, V.~E. 1973, Astrofizicheskie Issledovaniia Izvestiya
  Spetsial'noj Astrofizicheskoj Observatorii, 8, 3

\bibitem[{{Kolmogorov}(1933)}]{kol33}
{Kolmogorov}, A. 1933, {Giornale dell'Istituto Italiano degli Attuari}, 4, 83

\bibitem[{{Lane} {et~al.}(2012){Lane}, {Salinas}, \& {Richtler}}]{lan12}
{Lane}, R.~R., {Salinas}, R., \& {Richtler}, T. 2012, ArXiv e-prints

\bibitem[{{Majewski} {et~al.}(1999){Majewski}, {Siegel}, {Kunkel}, {Reid},
  {Johnston}, {Thompson}, {Landolt}, \& {Palma}}]{maj99}
{Majewski}, S.~R., {Siegel}, M.~H., {Kunkel}, W.~E., {et~al.} 1999, \aj, 118,
  1709

\bibitem[{{Mateo} {et~al.}(1996){Mateo}, {Mirabal}, {Udalski}, {Szymanski},
  {Kaluzny}, {Kubiak}, {Krzeminski}, \& {Stanek}}]{mat96}
{Mateo}, M., {Mirabal}, N., {Udalski}, A., {et~al.} 1996, \apjl, 458, L13

\bibitem[{{Niemi} {et~al.}(2010){Niemi}, {Hein{\"a}m{\"a}ki}, {Nurmi}, \&
  {Saar}}]{nie10}
{Niemi}, S.-M., {Hein{\"a}m{\"a}ki}, P., {Nurmi}, P., \& {Saar}, E. 2010,
  \mnras, 405, 477

\bibitem[{{Peng} {et~al.}(2008){Peng}, {Jord{\'a}n}, {C{\^o}t{\'e}},
  {Takamiya}, {West}, {Blakeslee}, {Chen}, {Ferrarese}, {Mei}, {Tonry}, \&
  {West}}]{pen08}
{Peng}, E.~W., {Jord{\'a}n}, A., {C{\^o}t{\'e}}, P., {et~al.} 2008, \apj, 681,
  197

\bibitem[{{Prugniel} \& {Heraudeau}(1998)}]{pru98}
{Prugniel}, P. \& {Heraudeau}, P. 1998, \aaps, 128, 299

\bibitem[{{Reda} {et~al.}(2004){Reda}, {Forbes}, {Beasley}, {O'Sullivan}, \&
  {Goudfrooij}}]{red04}
{Reda}, F.~M., {Forbes}, D.~A., {Beasley}, M.~A., {O'Sullivan}, E.~J., \&
  {Goudfrooij}, P. 2004, \mnras, 354, 851

\bibitem[{{Richtler}(2003)}]{ric03}
{Richtler}, T. 2003, in Lecture Notes in Physics, Berlin Springer Verlag, Vol.
  635, Stellar Candles for the Extragalactic Distance Scale, ed. D.~{Alloin} \&
  W.~{Gieren}, 281--305

\bibitem[{{Richtler}(2006)}]{ric06}
{Richtler}, T. 2006, Bulletin of the Astronomical Society of India, 34, 83

\bibitem[{{Richtler}(2012)}]{ric12a}
{Richtler}, T. 2012, ArXiv e-prints 1210.0045

\bibitem[{{Richtler} {et~al.}(2012){Richtler}, {Bassino}, {Dirsch}, \&
  {Kumar}}]{ric12b}
{Richtler}, T., {Bassino}, L.~P., {Dirsch}, B., \& {Kumar}, B. 2012, \aap, 543,
  A131

\bibitem[{{Rieke} \& {Lebofsky}(1985)}]{rie85}
{Rieke}, G.~H. \& {Lebofsky}, M.~J. 1985, \apj, 288, 618

\bibitem[{{Salinas} {et~al.}(2012){Salinas}, {Richtler}, {Bassino},
  {Romanowsky}, \& {Schuberth}}]{sal12}
{Salinas}, R., {Richtler}, T., {Bassino}, L.~P., {Romanowsky}, A.~J., \&
  {Schuberth}, Y. 2012, \aap, 538, A87

\bibitem[{{Sandage}(1978)}]{san78}
{Sandage}, A. 1978, \aj, 83, 904

\bibitem[{{Schlegel} {et~al.}(1998){Schlegel}, {Finkbeiner}, \&
  {Davis}}]{sch98}
{Schlegel}, D.~J., {Finkbeiner}, D.~P., \& {Davis}, M. 1998, \apj, 500, 525

\bibitem[{{Schuberth} {et~al.}(2010){Schuberth}, {Richtler}, {Hilker},
  {Dirsch}, {Bassino}, {Romanowsky}, \& {Infante}}]{sch10}
{Schuberth}, Y., {Richtler}, T., {Hilker}, M., {et~al.} 2010, \aap, 513, A52

\bibitem[{{Secker}(1995)}]{sec95}
{Secker}, J. 1995, \pasp, 107, 496

\bibitem[{{Slater} {et~al.}(2013){Slater}, {Bell}, {Schlafly}, {Juri{\'c}},
  {Martin}, {Rix}, {Bernard}, {Burgett}, {Chambers}, {Finkbeiner}, {Goldman},
  {Kaiser}, {Magnier}, {Morganson}, {Price}, \& {Tonry}}]{sla13}
{Slater}, C.~T., {Bell}, E.~F., {Schlafly}, E.~F., {et~al.} 2013, \apj, 762, 6

\bibitem[{{Smirnov}(1948)}]{smi48}
{Smirnov}, N. 1948, {The Annals of Mathematical Statistics}, 19, 279

\bibitem[{{Smith} {et~al.}(2004){Smith}, {Mart{\'{\i}}nez}, \&
  {Graham}}]{smi04}
{Smith}, R.~M., {Mart{\'{\i}}nez}, V.~J., \& {Graham}, M.~J. 2004, \apj, 617,
  1017

\bibitem[{{Sparks} {et~al.}(1985){Sparks}, {Wall}, {Thorne}, {Jorden}, {van
  Breda}, {Rudd}, \& {Jorgensen}}]{spa85}
{Sparks}, W.~B., {Wall}, J.~V., {Thorne}, D.~J., {et~al.} 1985, \mnras, 217, 87

\bibitem[{{Spitler} \& {Forbes}(2009)}]{spi09}
{Spitler}, L.~R. \& {Forbes}, D.~A. 2009, \mnras, 392, L1

\bibitem[{{Spitler} {et~al.}(2008){Spitler}, {Forbes}, {Strader}, {Brodie}, \&
  {Gallagher}}]{spi08}
{Spitler}, L.~R., {Forbes}, D.~A., {Strader}, J., {Brodie}, J.~P., \&
  {Gallagher}, J.~S. 2008, \mnras, 385, 361

\bibitem[{{Stocke} {et~al.}(2004){Stocke}, {Keeney}, {Lewis}, {Epps}, \&
  {Schild}}]{sto04}
{Stocke}, J.~T., {Keeney}, B.~A., {Lewis}, A.~D., {Epps}, H.~W., \& {Schild},
  R.~E. 2004, \aj, 127, 1336

\bibitem[{{Tal} {et~al.}(2009){Tal}, {van Dokkum}, {Nelan}, \&
  {Bezanson}}]{tal09}
{Tal}, T., {van Dokkum}, P.~G., {Nelan}, J., \& {Bezanson}, R. 2009, \aj, 138,
  1417

\bibitem[{{Temi} {et~al.}(2004){Temi}, {Brighenti}, {Mathews}, \&
  {Bregman}}]{tem04}
{Temi}, P., {Brighenti}, F., {Mathews}, W.~G., \& {Bregman}, J.~D. 2004, \apjs,
  151, 237

\bibitem[{{Tully}(1988)}]{tul88}
{Tully}, R.~B. 1988, {Nearby galaxies catalog}

\bibitem[{{van Dokkum} {et~al.}(2010){van Dokkum}, {Whitaker}, {Brammer},
  {Franx}, {Kriek}, {Labb{\'e}}, {Marchesini}, {Quadri}, {Bezanson},
  {Illingworth}, {Muzzin}, {Rudnick}, {Tal}, \& {Wake}}]{vdo10}
{van Dokkum}, P.~G., {Whitaker}, K.~E., {Brammer}, G., {et~al.} 2010, \apj,
  709, 1018

\bibitem[{{West}(1993)}]{wes93}
{West}, M.~J. 1993, \mnras, 265, 755

\bibitem[{{Yoon} {et~al.}(2006){Yoon}, {Yi}, \& {Lee}}]{yoo06}
{Yoon}, S.-J., {Yi}, S.~K., \& {Lee}, Y.-W. 2006, Science, 311, 1129

\end{thebibliography}

\end{document}